\documentclass[twocolumn,showpacs,
superscriptaddress,preprintnumbers,amsmath,amssymb]{revtex4}

\usepackage{graphicx}
\usepackage{dcolumn}
\usepackage{bm}

\def\Vec#1{\mbox{\boldmath$#1$}}
\def\Ren#1{\mbox{\boldmath$#1$}}

\begin{document}
  
  \preprint{}
  
  \title{
    Stable First-order Particle-frame Relativistic Hydrodynamics
    for Dissipative Systems
  }
  
  \author{Kyosuke Tsumura}
  \affiliation{
    Analysis Technology Center,
    Fujifilm Corporation,
    Kanagawa 250-0193, Japan
  }
  
  \author{Teiji Kunihiro}
  \affiliation{
    Yukawa Institute for Theoretical Physics, Kyoto University,
    Kyoto 606-8502, Japan
  }
  
  \begin{abstract}
    We propose a stable first-order relativistic dissipative hydrodynamic
equation in the particle frame
 (Eckart frame) 
for the first time.
 The equation  to be proposed was in fact previously  derived 
by the authors and a collaborator
from
   the
relativistic Boltzmann equation.
    We demonstrate 
that
 the equilibrium state is stable with respect to 
  the time evolution described  by our  hydrodynamic  equation
in the particle frame.
Our  equation may be a proper starting point
    for constructing second-order causal relativistic
    hydrodynamics, to replace Eckart's particle-flow theory.
  \end{abstract}
  
  \date{\today}
  \maketitle
  
  Relativistic hydrodynamics (RHD) is a useful tool
  for analyzing  slow and long wavelength behavior
  of relativistic many-particle systems
  in terms of  static and dynamic thermodynamic properties.
  In fact,  RHD is widely used
  in  astrophysics
  \cite{ast001}
  and the phenomenology of relativistic heavy ion collisions
  \cite{qcd001}.
 Since works demonstrating the success of  perfect hydrodynamics
in describing the phenomenology of the Relativistic Heavy Ion Collider (RHIC) at BNL
  \cite{qcd001, qcd002, qcd003},
  we are witnessing  a growing interest
  in  RHD for  {\em dissipative} systems
  \cite{muronga07,chaudhuri,romatschke,biro07}.
  Indeed, 
there have been many works attempting  to show how small can be the transport 
coefficients  of  strongly-interacting systems
  composed of hadrons or quarks and gluons,
with many of these employing the so-called AdS/CFT
correspondence hypothesis
  \cite{qcd012}.
It should be noticed, however, that
  the theory of RHD for  dissipative systems
is not 
clearly established,
 although there 
have been many fundamental studies
  since Eckart's pioneering work
  \cite{hen001}.
  
  We 
 identify the following three fundamental problems 
regarding  relativistic hydrodynamic equations (RHDEs)
  for  dissipative fluids
  \cite{TKO}:
  (a)~ambiguities in the forms of the equations
  \cite{hen001,hen002,hyd001,mic005,muronga07,romatschke};
  (b)~the unphysical instability of the equilibrium state
  in the theory of the so-called first-order equations,
in particular in the Eckart frame
  \cite{hyd002},
 defined below; (c)~the lack of causality in the first-order equations
  \cite{app002, mic001, mic004, mic005}.
  The present paper is concerned with the first two problems.
  Although the unphysical instability
of the equilibrium state
may be attributable to the lack of causality,
  and the Israel-Stewart equations
  with second-order time-derivative are 
presently being examined
in connection to this problem
  \cite{mic004,muronga07,chaudhuri,romatschke},
  we 
emphasize that
  the first two problems and the third one have different origins,
  and the first two must be resolved
  before the third 
 is addressed.
Note that  the causality problem also exists
 in non-relativistic cases
  and is 
in essence
a problem of
 how to incorporate
  the 
space-time scales shorter than those
  corresponding to the mean-free path,
  beyond those in the usual hydrodynamic regime.
  We also remark that
the proper form of Israel-Stewart-type equations
has not yet been definitely determined 
  \cite{muronga07,romatschke}.
  
  Let us 
represent  the flow velocity
  by $u^{\mu}$, with $u_{\mu} \, u^{\mu} = g^{\mu\nu} \, u_{\mu} \, u_{\nu} = 1$
  ($g^{\mu\nu} = {\rm diag}(+1, -1, -1, -1)$).
  In the relativistic theory,
  the rest frame of the fluid 
and  the flow velocity $u^{\mu}$
  cannot be uniquely defined
  when 
there exist 
viscosity and heat conduction.
  In the phenomenological theories
  \cite{hen001, hen002},
  the ambiguity of the flow velocity $u^{\mu}$
is resolved by 
placing constraints on the dissipative part
  of the energy-momentum tensor, $\delta T^{\mu\nu}$,
  and the particle current, $\delta N^{\mu}$.
  Landau and Lifshitz defined $u^{\mu}$ 
such  that
  there is no dissipative energy density, energy flow nor particle density;
  i.e.,
we have the constraints
$\delta T^{\mu\nu} \, u_{\nu} = 0$ 
(referred to as ET)
and 
$u_{\mu} \, \delta N^{\mu} = 0$~ (EN).
  This frame is called the energy frame.
 Contrastingly,
Eckart chose the particle frame,
in which there is no dissipative contribution to the particle current;
  i.e.,
we have
$\delta N^{\mu} = 0$ (PN), together with 
$u_{\mu} \, u_{\nu} \, \delta T^{\mu\nu} = 0$ (PT):
These conditions imply that there is no dissipative contribution to
  the energy density in this frame.
However, it should be noted that
the seemingly plausible constraint PT on $\delta T^{\mu\nu}$ is
problematic, as
shown in
  \cite{TKO}
  and 
explained below.
  
  Recently,
  Tsumura, Kunihiro (the present authors) and Ohnishi (abbreviated as TKO)
  \cite{TKO}
  derived generic covariant hydrodynamic equations for a viscous fluid
through a reduction of 
the 
dynamics
described by the  relativistic Boltzmann equation
  in a 
systematic manner, with no heuristic arguments,
  on the basis of the so-called renormalization group (RG) method
  \cite{rgm001,env001,HK02}.
  This was 
 done  by introducing the macroscopic frame vector $\Ren{a}^{\mu}$
that defines the macroscopic Lorenz frame,
  in which the slow dynamics 
are described.
  The generic equation 
derived by TKO can produce
a relativistic dissipative hydrodynamic equation
  in any frame with 
the appropriate choice of  $\Ren{a}^{\mu}$;
  the resulting equation in the energy frame
  coincides with that of Landau and Lifshitz
  \cite{hen002},
  while that in the particle frame
  is similar to, but 
slightly
 different from, the Eckart equation.
 Interestingly, 
 the TKO equation in the particle frame
  does not satisfy the constraints PT on $\delta T^{\mu\nu}$
  but, 
instead,
 satisfies 
$\delta T^{\mu}_{\,\,\,\mu} = 0$, 
which we call PT',
 together with PN.
  It should be 
noted that
  the new constraints, PT',
are identical to a matching condition
  postulated by Marle and Stewart (MS)
  in 
the derivation of the RHD
  from the Boltzmann equation with 
use of 
 Grad's moment theory
  \cite{mic003}.
  We call the constraints PT', together with PN,
  the Grad-Marle-Stewart (GMS) constraints.
  In 
\cite{TKO}, TKO proved that
  the simultaneous constraints PT and PN cannot be compatible
  with the underlying Boltzmann equation
  if the hydrodynamic equation describes
  the slow,
long wavelength 
limit of the solutions of the Boltzmann equation.
This is interesting in connection to problem (b), i.e.,
the fact that the solutions of the Eckart equation
  around the thermal equilibrium 
are unstable
  \cite{hyd002},
  while the Landau theory is stable.
  
An immediate question is whether
  the solutions of the new equations in the particle frame are stable
  around the thermal equilibrium.
  In fact, the
hydrodynamic equations of MS and TKO
  in the particle frame are of different forms,
  although both 
 satisfy the constraints PT' and PN.
  In the present paper,
  we examine the stability problem 
for the new equations in the particle frame.
Because  second-order equations, such as
 the Israel-Stewart equations,
  are usually constructed in the particle frame,
  as an extension of the Eckart equation,
  finding a stable first-order equation in the particle frame
  is 
of fundamental significance.
As the RG method 
has been employed to construct the slow dynamics
of various 
systems 
through the explicit construction
of  the slow,
  stable manifold of the dynamics,
we conjecture that  the hydrodynamic equation obtained
  as the slow, long wavelength limit of the Boltzmann equation
  on the basis of the RG method
will provide a description in which 
the thermal equilibrium state is stable.
  We 
demonstrate that
  this is indeed the case
  by performing
a linear stability analysis
using the EOS and the transport coefficients for a rarefied gas.
By contrast, we find that
 the MS equation,
like the Eckart equation,
 is 
 unstable.
Hence, for the first time, a
stable RHDE 
  is obtained in the particle frame.
We believe that this will provide a 
sound starting point
  for the construction of the proper second-order equations.
  
  The energy-momentum tensor for our equation in the particle frame reads
  \begin{eqnarray}
    T^{\mu\nu} &=& \epsilon \, u^{\mu} \, u^{\nu} - p \,
    \Delta^{\mu\nu}
    + \lambda \, u^{\mu} \, \nabla^{\nu} T + \lambda \, u^{\nu} \, 
    \nabla^{\mu} T\nonumber\\
    & & {} + \zeta \, (3 \, u^{\mu} \, u^{\nu} - \Delta^{\mu\nu}) \,
	  [ - ( 3 \, \gamma - 4)^{\mbox{-}2} \, \nabla \cdot u ]\nonumber\\
	  & & {} + \eta \, ( \nabla^{\mu} u^{\nu} +
	  \nabla^{\nu} u^{\mu} - 2/3 \, \Delta^{\mu\nu} \, \nabla \cdot
	  u ),
  \end{eqnarray}
  while the particle current 
is given by $N^{\mu} = n \, u^{\mu}$,
  with $\Delta^{\mu\nu} \equiv g^{\mu\nu} - u^\mu \, u^\nu$
  and $\nabla^{\mu} \equiv \Delta^{\mu\nu} \, \partial_{\nu}$.
  Here $T$, $\mu$, $\epsilon$, $p$, $n$ and $\gamma$ are
  the temperature,
  the chemical potential,
  the internal energy,
  the pressure,
  the particle density and the ratio of the specific heats, respectively,
  and  $\zeta$, $\lambda$ and $\eta$ denote
  the bulk viscosity, the heat conductivity and the shear viscosity,
  respectively.
  The MS equations are obtained from the above equations
through  the replacements
  $-\zeta ( 3 \, \gamma - 4)^{\mbox{-}2} \, \nabla \cdot u \longrightarrow
  + \zeta ( 3 \, \gamma - 4)^{\mbox{-}1} \, \nabla \cdot u$ and
  $\lambda \nabla^{\mu} T \longrightarrow \lambda(\nabla^{\mu} T - T \, Du^{\mu})$,
  where $D \equiv u^{\nu} \, \partial_{\nu}$.
  One can easily check that both equations satisfy the GMS constraints.
  Nevertheless,
 we
find  the following differences between them:
  (A)~
the thermal forces in the MS equations contain 
  the time-like derivative of the flow velocity
$Du^{\mu}$,
  while those 
in our equations
 involve only the space-like derivative $\nabla^{\mu}$, and
  (B)~the sign of the thermodynamic force owing to 
the bulk viscosity 
in our equation
  is the same 
 as that in the Landau equation
 and opposite 
that in the MS equation.
  We can trace 
the two characteristic features of our theory
back   to the simple ansatz that
only  the spatial inhomogeneity, over distances
  of the order of the mean free path, 
 is the origin of the dissipation.
  It should be 
noted that
  the same ansatz for the non-relativistic case
leads naturally to the Navier-Stokes equation,
 as shown in
  \cite{HK02},
  and hence our framework can be interpreted
  as the most natural covariantization of the non-relativistic case.
  
  The thermal equilibrium state is 
given  by
  $u^{\mu}(x) = (1,\,0,\,0,\,0) \equiv u_0^{\mu}$,
  $T(x) = T_0$ and ${\mu}(x) = \mu_0$,
  with $T_0$ and $\mu_0$ being constant.
This is a trivial solution to the equations.
  Let us investigate the linear stability of the equilibrium solution.
  Writing $T(x) = T_0 + \delta T(x)$,
  $\mu(x) = \mu_0 + \delta \mu(x)$ and $u^{\mu}(x) = u_0^{\mu} + \delta u^{\mu}(x)$,
  we examine the time evolution of the deviations
  in the linear approximation
using the evolution equation
given by
  $\partial_{\mu} T^{\mu\nu} =0$ and $\partial_{\mu} N^{\mu} = 0$.
  Here we note that
  the independent 
variables are 
 the
five
 quantities $\delta T(x)$, $\delta \mu(x)$ and $\delta u^i(x)$ ($i = 1,\,2,\,3$),
  because $\delta u^0(x) = 0$, due to the constraint
  $u_{\mu}(x) \, u^{\mu}(x) = 1$.
  
  In terms of the Fourier components
  $\tilde{\Phi}_\alpha(k) \equiv {}^\mathrm{t}(\delta \tilde{u}^1(k),\, 
  \delta \tilde{u}^2(k),\,
  \delta \tilde{u}^3(k),\, \delta \tilde{T}(k),\, \delta
  \tilde{\mu}(k))$, defined through
  $\Phi_{\alpha}(x) = \int \!\! \frac{\mathrm{d}^4k}{(2\pi)^4} \, 
  \tilde{\Phi}_{\alpha}(k) \, \mathrm{e}^{\mbox{-}i k \cdot x}$,
  the linearized hydrodynamic equation
reduces to
the  algebraic equation
  $\sum_{\beta=1}^5 \, M_{\alpha\beta} \, \tilde{\Phi}_\beta = 0$, 
  with
  \begin{eqnarray}
    \label{eq:2-003}
    M_{\alpha\beta} \equiv \left(
    \begin{array}{ccccc}
      \mathcal{L}_1 &
      0 &
      0 &
      0 &
      0\\
      0 &
      \mathcal{L}_1 &
      0 &
      0 &
      0\\
      0 &
      0 &
      \mathcal{L}_1 - \mathcal{L}_2 \, (k^3)^2 &
      i \, \mathcal{L}_3 \, k^3 &
      i \, \mathcal{L}_4 \, k^3\\
      0 &
      0 &
      -i \, \mathcal{L}_5 \, k^3 &
      \mathcal{L}_6 &
      \mathcal{L}_7\\
      0 &
      0 &
      -i \, \mathcal{L}_8 \, k^3 &
      \mathcal{L}_9 &
      \mathcal{L}_{10}
    \end{array}
    \right),
  \end{eqnarray}
  where we have set $k^{\mu} = (k^0,\,0,\,0,\,k^3)$
  without
loss of generality.
The first and second components of $\tilde{\Phi}_{\alpha}$ describe
  the transverse mode,
  while the third component the longitudinal one.
  Here $\mathcal{L}_{i=1 \sim 10}$ are given by
  $\mathcal{L}_1 \equiv (\epsilon + p) \, (-i \, k^0) +
  \eta \, |\Vec{k}|^2$,
  $\mathcal{L}_2 \equiv - \eta /3 - \zeta_P$,
  $\mathcal{L}_3 \equiv {\partial p}/{\partial T}
  - \lambda \, (-i \, k^0)$,
  $\mathcal{L}_4 \equiv {\partial p}/{\partial \mu}$,
  $\mathcal{L}_5 \equiv - (\epsilon + p) + 3 \, \zeta_P\, (-i \, k^0)$
  $\mathcal{L}_6 \equiv {\partial \epsilon}/{\partial T} \, (-i \, k^0) + \lambda \, |\Vec{k}|^2$,
  $\mathcal{L}_7 \equiv {\partial \epsilon}/{\partial \mu} \, (-i \, k^0)$,
  $\mathcal{L}_8 \equiv - n$,
  $\mathcal{L}_9 \equiv {\partial n}/{\partial T} \, (-i \, k^0)$
  and $\mathcal{L}_{10} \equiv {\partial n}/{\partial \mu} \, (- i \, k^0)$,
  with $\zeta_P \equiv \zeta \, (3  \, \gamma - 4)^{\mbox{-}2}$
  being the effective bulk viscosity in the particle frame.
In the above, the quantities,
$\epsilon$,
  $p$,
  $n$,
  $\gamma$,
  $\zeta$,
  $\lambda$,
  $\eta$,
  ${\partial \epsilon}/{\partial T}$,
  ${\partial \epsilon}/{\partial \mu}$,
  ${\partial p}/{\partial T}$,
  ${\partial p}/{\partial \mu}$,
  ${\partial n}/{\partial T}$
  and ${\partial n}/{\partial \mu}$
take their  
 equilibrium values,
with $T = T_0$ and $\mu = \mu_0$.
  
  The existence condition of 
a solution reads $\det M = 0$,
  which 
reduces to
  \begin{eqnarray}
    \label{eq:2-004}
    & & \mathcal{L}^2_1 \, \Big[
      ( \mathcal{L}_1 - |\Vec{k}|^2 \, \mathcal{L}_2 )
      \,
      ( \mathcal{L}_6 \, \mathcal{L}_{10} - \mathcal{L}_7 \, \mathcal{L}_9 )
      - |\Vec{k}|^2 \, \mathcal{L}_5
      \,
      ( \mathcal{L}_3 \, \mathcal{L}_{10}\nonumber\\
    & & {}- \mathcal{L}_4 \, \mathcal{L}_9 )
      - |\Vec{k}|^2 \, \mathcal{L}_8
      \,
      ( \mathcal{L}_4 \, \mathcal{L}_6 - \mathcal{L}_3 \, \mathcal{L}_7 )
      \Big] = 0.
  \end{eqnarray}
  This equation gives the dispersion relation $k^0 = k^0(|\Vec{k}|)$
  for the hydrodynamic modes,
  and the stability condition for the equilibrium state
  reads $\mathrm{Im}k^0 \le 0$,  $\forall \, |\Vec{k}|$.
  
  We see the dispersion relation for the transverse mode
  is given by $\mathcal{L}_1 = 0$,
whose solution 
 is  $k^0 = - i \, \eta\, |\Vec{k}|^2/(\epsilon + p)$.
  Thus, 
we find that the transverse mode is stable.
  
  Here we again stress that
  the equation we study does not contain a term
  proportional to $D u^{\mu}$ in the thermal force for the heat flow.
  What would happen if such a term 
were present
  in the thermal forces, as in the 
case of the MS 
and the Eckart theories?
  In this case, the corresponding equation becomes 
  $\mathcal{L}_1 =(\epsilon + p) \, (-i \, k^0)
  - T \, \lambda \, (-i \, k^0)^2 + \eta \, |\Vec{k}|^2 = 0$,
  which 
possesses a root with a positive imaginary part,
  and hence an unstable transverse mode appears.
We emphasize that
  this instability is inevitable
if the heat flow term contains $D u^{\mu}$ \cite{biro07}.
  
  Next, we examine the dispersion relations of the longitudinal modes.
  We first 
consider the simple but
interesting case
in which the heat conductivity 
vanishes (i.e., $\lambda = 0$),
  but the bulk and the shear viscosities may be positive
(i.e.,   $\zeta \neq 0$ and $\eta \neq 0$).
This simple 
case is often
studied in the literature.
  We 
subsequently carry out 
a full analysis 
in which  all the transport coefficients,
  including $\lambda$, may be positive.
  
  In the simple case with $\lambda = 0$,
  the equation has
the
root $k^0 = 0$ and those
satisfying   $a_0 \, (-i \, k^0)^2 +  b_0 \, (-i \, k^0) + c_0 = 0$,
  with
  $a_0 \equiv (\epsilon + p)\,  \{ \epsilon \,,\, n \}$,
  $b_0 \equiv |\Vec{k}|^2 \, [({4\eta}/{3} + \zeta_P) \, \{\epsilon
    \,,\, n\} - 3 \,\zeta_P \, \{ p \,,\, n \}]$
  and
  $c_0 \equiv |\Vec{k}|^2 \, [(\epsilon + p) \, \{ p \,,\, n \} + n \, \{\epsilon \,,\, p \}]$,
  where we have written
  the Jacobian as  $\{ F \,,\, G \} \equiv \partial(F\,,\, G)/\partial(T\,,\, \mu)$.
  Now, 
a simple analysis of the algebraic equation
  $a_0 \, \omega^2 + b_0 \, \omega + c_0 = 0$ with $\omega = -i \, k^0$
  shows that the necessary and sufficient condition for $\exists \, k^0$
  with $\mathrm{Im}k^0 \le 0$ is that $b_0/a_0 \ge 0$ and $c_0/a_0 \ge 0$.
  Owing to the properties of the Jacobian and the 
thermodynamic relations,
  the last inequality generally holds,  because the l.h.s 
can be rewritten as
  $c_0/a_0 = |\Vec{k}|^2 \, (\partial p/\partial \epsilon)_S = |\Vec{k}|^2 \, c_s^2$,
  with $c_s$ being the sound velocity.
  Then, the stability condition 
reduces to $b_0/a_0 \ge 0$,
  which is equivalent to
  \begin{eqnarray}
    \label{eq:2-008}
    {4\eta}/{3} + \zeta_P \, [ 1 - 3 \, (\partial p/{\partial \epsilon})_n ] \ge 0.
  \end{eqnarray}
  This is a new condition 
 that involves  not only the EOS but also the bulk and shear viscosities.
  It can be shown analytically
  \cite{nex001}
  that this inequality is satisfied
  at least for a rarefied gas in the massless limit.
To see this, first notice that $\epsilon = 3 \, p$
  for 
a relativistic gas composed of massless particles.
  Then the inequality 
reduces to 
the trivial one $\eta \ge 0$,
  because the second term with a bracket on the l.h.s vanishes,
  provided that the effective bulk viscosity,
  $\zeta_P = \zeta \, (3 \, \gamma -4)^{-2}$,
 is finite in the massless limit.
  In fact, it can be shown that this is the case
  using the microscopic formula for $\zeta$
  \cite{nex001},
  although $3 \, \gamma - 4 \rightarrow 0$ in the massless limit.
  We also remark that 
 numerical calculations
  using the viscosities $\zeta$ and $\eta$ obtained from the Boltzmann equation
reveal  that the inequality (\ref{eq:2-008}) is always satisfied,
  even for a rarefied gas of massive particles.
  Instead of presenting the numerical results for this limiting case,
  we 
present the results
  for the 
general  case,
i.e., that in which $\lambda\not= 0$,
$\zeta\not= 0$ and $\eta \not=0$,
below.
  
  We now demonstrate that
the thermal equilibrium state
is stable with respect to the dynamics
described by our equation, 
  even when the heat conductivity, $\lambda $, is finite.
  The dispersion equation for the longitudinal modes 
 is obtained 
from the roots of the 
cubic equation
  $a \, \omega^3 + b \, \omega^2 + c \, \omega + d = 0$,
 with $\omega =-i \, k^0$,
  where the coefficients are given by
  $a \equiv a_0 + |\Vec{k}|^2
  \, 3 \,  \zeta_P \, \lambda \, ({\partial n}/{\partial \mu})_T$,
  $b \equiv  b_0 + n \, \lambda \, |\Vec{k}|^2 \, ({\partial \epsilon}/{\partial \mu})_T$,
  $c \equiv c_0 + |\Vec{k}|^4 \, ({4\eta}/{3} +  \zeta_P) \, \lambda \,
  ({\partial n}/{\partial \mu})_T$ and
  $d \equiv |\Vec{k}|^4 \, n \,  \lambda \, ({\partial p}/{\partial \mu})_T$.
  The condition ${\rm Im} k^0 \le 0$ implies 
that the above equation for
 $\omega$ 
has roots only in the left half plane or on the
  imaginary axis in the complex $\omega$ plane.
 An elementary analysis shows that this condition is given by
  \begin{eqnarray}
    \label{eq:022}
    a \ge 0, \, b \ge 0, \, d \ge 0 \,\, {\rm and} \,\, b \, c - a \, d \ge 0.
  \end{eqnarray}
  Here, the equality
holds in  the case that the imaginary part of $k^0$ vanishes.
Note that the above equalities imply that $c \ge 0$.
  
\begin{widetext}
  
    \begin{figure}[tHb]
      \begin{center}
	\begin{minipage}{.30\linewidth}
          \includegraphics[width=\linewidth]{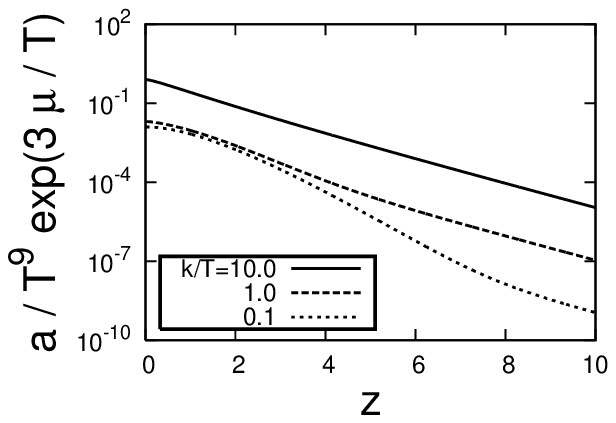}
	\end{minipage}
	\begin{minipage}{.30\linewidth}
          \includegraphics[width=\linewidth]{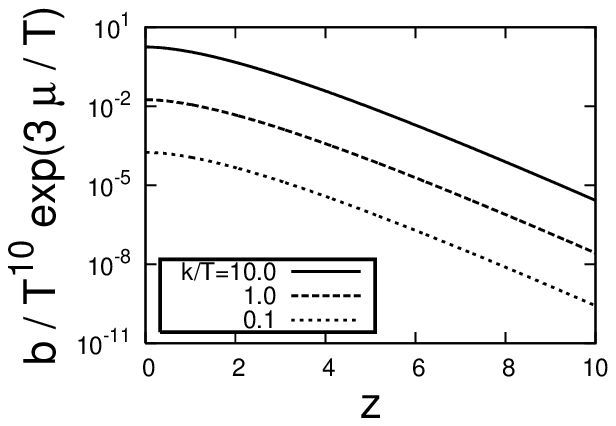}
	\end{minipage}
	\begin{minipage}{.30\linewidth}
          \includegraphics[width=\linewidth]{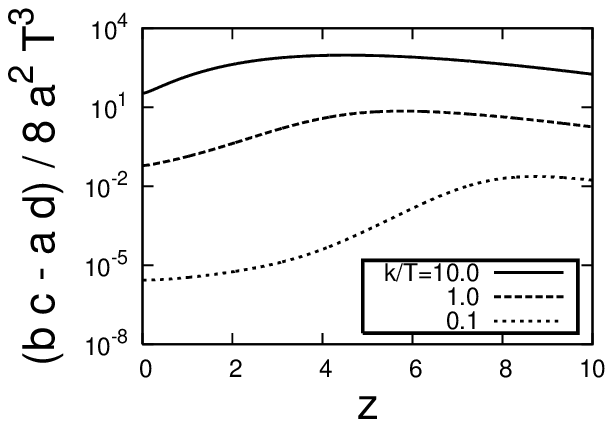}
	\end{minipage}
      \end{center}
      \caption{
	The $z = m/T$ dependence of 
	$a/( T^{9} \, \mathrm{e}^{3\mu/T})$,
	$b/( T^{10} \, \mathrm{e}^{3\mu/T})$
	and
	$(b \, c - a \, d)/({8 \, a^2 \, T^3})$
	for $|\Vec{k}|/T=0.1, \, 1, \, 10$.
	Some factors have been multiplied so that
	the variables become independent of $\mu$.
	It is seen that all these quantities are positive in all cases.
      }
      \label{fig:002}
    \end{figure}
    
\end{widetext}

  Now
we demonstrate that
  these inequalities are satisfied for rarefied systems.
  For
a relativistic free gas,
  we have
  $n = (2\pi)^{\mbox{-}3} \, 4 \, \pi \, m^3 \,
  \mathrm{e}^{\frac{\mu}{T}} \, [ z^{\mbox{-}1} \, K_2(z) ]$,
  $\epsilon = m \, n \, [ K_3(z)/K_2(z) - z^{\mbox{-}1} ]$,
  $p = n \, T$ and
  $\gamma = 1 + [ z^2 + 3 \, \hat{h} - (\hat{h} - 1)^2 ]^{\mbox{-}1}$
  with $z = m/T$ and $\hat{h} = (\epsilon + p) / n \, T$
  being the reduced enthalpy.
  Here, $K_2(z)$ and $K_3(z)$ denote the second and third modified
  Bessel functions, respectively.
It is seen that the positivity condition $d > 0$ holds
  from the formula $p = n \, T$ with $n \propto \mathrm{e}^{\frac{\mu}{T}}$,
  which implies that $(\partial p/\partial \mu)_T > 0$.
It remains to demonstrate the rest of the inequalities,
  i.e., $a \ge 0$, $b \ge 0$ and $b \, c - a \, d \ge 0$,
  for which we need explicit forms of the transport coefficients as well.
  The transport coefficients $\zeta$, $\lambda$ and $\eta$ for 
a rarefied gas
  can be obtained from the collision term in the Boltzmann equation.
  The Galerkin approximation 
using the Ritz polynomial expansion
  \cite{mic001}
  with a constant cross section $\sigma$ 
in  the collision integral gives
  $\zeta = \frac{1}{32 \pi} \frac{T}{\sigma}
  \mathrm{e}^{\mu/T} \big[z^2
    K^2_2(z) [ (5 - 3 \gamma) \hat{h} - 3 \gamma ]^2\big]/
	 [2 K_2(2 z) + z K_3(2 z)]$,
  $\lambda = \frac{3}{32 \pi} \frac{1}{\sigma}
	 \mathrm{e}^{\mu/T} \big[z^2 K^2_2(z) [ \gamma / (\gamma - 1)]^2\big]
	 /[(z^2 + 2) K_2(2 z) + 5 z K_3(2 z)]$ and
  $\eta = \frac{15}{32 \pi} \frac{T}{\sigma}
	 \mathrm{e}^{\mu/T} [z^2 K^2_2(z) \hat{h}^2]/
		[(5 z^2 + 2) K_2(2 z) + (3 z^3 + 49 z)
		  K_3(2 z)]$.
Note that
  all the transport coefficients are proportional
  to the inverse of the cross section, $\sigma$.
This implies that a strongly (weakly) interacting system has
  small (large) transport coefficients.
  The numerical results for $a$, $b$ and $b \, c - a \, d$ are
displayed in Fig.{\ref{fig:002}},
  where the $z = m/T$ dependence 
  is shown 
 using $\sigma \, T^2 = 1$
  for a wide range of
values of 
the three momentum:
$|\Vec{k}|/T=0.1$ - $10$.
We have 
confirmed that
  the positivity of these quantities holds
  for a wide range of values of the cross section: $\sigma \, T^2 = 0.01$ - $10$.
 We 
point out that a rarefied gas is a system
in which   dissipative effects are most significant.
  Thus,
we have demonstrated that
  the thermal equilibrium solution is stable
within the description provided by our hydrodynamic
 equation in the particle frame.
Obviously,  
a solution with flow is unstable in a viscous fluid, as
it must
 relax
to the
equilibrium state.

  In this Letter,
  we first 
pointed out  that
  the  constraint proposed by Eckart,
$u_{\mu} \, u_{\nu} \, \delta T^{\mu\nu} = 0$ (PT),
is 
incompatible
  with the underlying relativistic Boltzmann equation.
This important point  has been
largely unnoticed.
  We 
then showed that the 
reduction of
  the Boltzmann equation
employing the RG method
leads to 
an RHDE
  in the particle frame,
that satisfies the constraint $\delta T^{\mu}_{\,\,\,\mu} = 0$,
  while
  $u_{\mu} \, u_{\nu} \, T^{\mu\nu} = \epsilon - 3 \, \zeta_P \, \nabla \cdot u$,
  which includes a contribution from the flow as well as the internal energy.
  This equation might imply that the energy density of an expanding system
  extracted from the hydrodynamic analysis can be erroneous.
  We have demonstrated that
  the solution around the equilibrium state
  in the new equation
  is stable.
This was done by carrying out a linear stability analysis
using   the EOS and the transport coefficients for a rarefied gas.
We conclude that Eq.~(1) 
represents the first viable possibility as a stable,
first-order, particle-frame RHDE for a viscous
fluid.
This is significant
because  the Israel-Stewart causal equation is
  usually constructed in the particle frame with
  PT.
  A detailed 
presentation  of 
this work
  and applications of the new equation studied here
will be reported elsewhere.
  
  We are grateful to Dirk Rischke and Miklos Gyulassy for comments.
  T. K. is supported by a Grant-in-Aid
  for Scientific Research by Monbu-Kagakusyo of Japan
  (No. 17540250), by  the 21st Century COE 
  ``Center for Diversity and Universality in Physics" of Kyoto
  University and by YIPQS at the 
Yukawa Institute for Theoretical Physics, Kyoto.
  
  \newcommand{\btem}{\bibitem}

\end{document}